\documentclass[useAMS,usenatbib,usegraphicx]{mn2e}
\usepackage{epsfig}
\usepackage{graphics}
\usepackage{color}
\usepackage{amsmath}
\usepackage{graphicx,dblfloatfix}
\usepackage{aas_macros}
\def\kms{\mbox{${\rm km~s}^{-1}$}}

\def\aj     {{\it Astron.\nobreak\ J.,}\nobreak\ }

\def\apj    {{\it Astro\-phys.\nobreak\ J.,}\nobreak\ }
\def\apjs   {{\it Astro\-phys.\nobreak\ J.\ Suppl.,}\nobreak\ }

\def\pasp   {{\it Publ.\ astr.\ Soc.\ Pacif.,}\nobreak\ }

\def\name#1 {{\it #1\/}}
\def\vol#1  {{\bf #1\/}}
\title[Stellar Populations of Shell Galaxies] {Stellar Populations of Shell Galaxies}

\author[S. Carlsten, G. K. T. Hau, \& A. Zenteno]  
{S. Carlsten,$^{1}$\thanks{sgc1@rice.edu} G. K. T. Hau,$^{2}$ A. Zenteno$^{3}$  \\ 
$^1$Rice University, 6100 Main Street, Houston, TX, 77005, USA\\
$^2$European Southern Observatory, Alonso de Cordova 3107, Chile \\
$^3$Cerro Tololo Inter-American Observatory, Casilla 603, La Serena, Chile\\
} 
\date{\today} \pubyear{2016}

\begin{document}

\def\zsun{\mbox{${\rm Z}_{\odot}$}}
\def\square{\vrule height 4.5pt width 4pt depth -0.5pt}
\def\threesquares{\square~\square~\square\ }
\def\remark#1{{\threesquares\tt#1~\threesquares}}

\newcommand{\mean}[1]{\mbox{$<$#1$>$}}
\newcommand{\h}[1]{\mbox{$h_#1$}}
\newcommand{\hbeta}{\mbox{H${\beta}$}}
\newcommand{\mg}[1]{\mbox{Mg$_#1$}}
\newcommand{\mgb}{\mbox{Mg$\ b$}}
\newcommand{\fe}[1]{\mbox{Fe$_{#1}$}}


\maketitle
\begin{abstract}
We present a study of the inner (out to $\sim$1 R$_{\mathrm{eff}}$) stellar populations of 9 shell galaxies. We derive stellar population parameters from long slit spectra by both analyzing the Lick indices of the galaxies and by fitting Single Stellar Population model spectra to the full galaxy spectra. The results from the two methods agree reasonably well. Many of the shell galaxies in our sample appear to have lower central $\mathrm{Mg}_{2}$ index values than non-shell galaxies of the same central velocity dispersion, which is likely due to a past interaction event. Our shell galaxy sample shows a relation between central metallicity and velocity dispersion that is consistent with previous samples of non-shell galaxies.
Analyzing the metallicity gradients in our sample, we find an average metallicity gradient of -0.16$\pm$0.10 dex per decade in radius. We compare this with formation models to constrain the merging history of shell galaxies. We argue that our galaxies likely have undergone major mergers in their past but it is unclear whether the shells formed from these events or from separate minor mergers. Additionally, we find evidence for young stellar populations ranging in age from 500 Myr to 4--5 Gyr in four of the galaxies, allowing us to speculate on the age of the shells. For NGC 5670, we use a simple dynamical model to find the time required to produce the observed distribution of shells to be consistent with the age of the young subpopulation, suggesting that the shells likely formed from the same event that led to the young subpopulation.
\end{abstract}


\section{Introduction} \label{sec:introduction} 

Shell galaxies are early type galaxies that have faint stellar shell-like structures surrounding them. First noticed by \citet{arp66} and catalogued by \citet{malin83}, shell galaxies have since been studied as products of galactic interaction and mergers. 
An understanding of shell galaxies might lead to an understanding of merging history, potential shape, and dark matter distribution for other early type galaxies \citep{ebrova13}. 
Because of the relative ubiquitousness of shell galaxies (\citet{malin83} estimate $\sim$10{\%} of isolated early type galaxies possess shells, while \citet{kraj11} finds $\sim$3.5{\%} of the 260 galaxy ATLAS$^{\rm 3D}$ sample have shells), they represent an important tool in understanding galaxy evolution. 

Since their discovery, many mechanisms have been proposed for the formation of shells. The most widely accepted mechanisms today fall into two categories: weak interaction and merger models.  The weak interaction model was first proposed by \citet{thomson90} and extended by \citet{thomson91}. The model posits that a fly-by interaction between a small satellite galaxy and a larger galaxy could excite shell-like density waves in the thick disk population of the larger galaxy. The type of shell galaxy produced \citep{prieur90} would depend on the viewing angle of the parent galaxy. The flyby would likely not change the stellar population properties or kinematics of the parent galaxy. However, \citet{hau94} found that a high speed flyby can spin up the outer layers of the host galaxy resulting in a kinematically decoupled core (KDC). The simulations by \citet{thomson90} predict shell lifetimes of 10Gyr whereas the later simulations by \citet{thomson91} predict lifetimes of only 3Gyr. 

The second principal model of shell formation is the idea that shells form from a galactic merger. The lack of nearby satellite galaxies around Fornax A led \citet{schweizer80} to conclude that the shells and the inclined disk of ionized gas were caused by an infalling galaxy. \citet{quinn84} took the idea of mergers and proposed that shells could form through ``phase-wrapping". He performed simple N-body simulations of a small disk galaxy falling into the potential of a host galaxy. The accreted stars oscillate back and forth forming shells. Essentially, shells form where the accreted stars reach the turn-around points of their orbits. These are the points where the stars are moving the slowest and therefore their density increases and quasi-static shells are observed. The shells would eventually degrade due to the velocity dispersion in the accreted stars. \citet{quinn84} cites 1Gyr as an average lifetime for the shell structures to last. Later on, others showed that elliptical accreted galaxies could also form shells through phase-wrapping \citep{dupraz86}. 

The early simulations showed that shells could form from the accretion of a satellite galaxy that was only a fraction of the mass of the primary \citep[10:1 or greater mass ratios were used in][]{quinn84}. Others have shown that shells could form from a major merger of equal mass galaxies. For example, \citet{hernquist92} showed that a merger of two equal sized disk galaxies could produce shells. Furthermore, in their simulation, shells were aligned with the major axis of the elliptical and were present over a large range in radii. Both of these features match observations and are somewhat problematic for the minor merger model \citep{ebrova13}. Later simulations by \citet{gonzalez05} demonstrate that shells can form from 3:1 and 2:1 disk mergers. Interestingly, they found that shells were not formed from 1:1 disk mergers, possibly indicating that the shells \citet{hernquist92} found were due to their specific (and improbable) initial orbital geometry whereby the planes of the disk galaxies initially coincided with the orbital plane. \citet{gonzalez05} found that shells were most prominent in mergers of disk galaxies without an initial bulge component. They concluded, however, that these mergers were less likely to be the source of real ellipticals because of the results' insufficient central densities and triaxial shape. With that said, several authors have still argued for a major merger cause for certain shell galaxies. For instance, \citet{serra06}, argue for a major merger origin for the shells in IC 4200 based on the presence of a massive, warped HI disk and a young central stellar population. Additionally, \citet{goudfrooij01}, argue for a major merger for the creation of shell galaxy Fornax A due to the presence of molecular gas and a young subpopulation of globular clusters. 

Distinguishing between a major merger and a minor merger may be possible through the metallicity gradient present in the galaxy. Formation of massive ellipticals consists of two ``phases" \citep{oser10,feldmann11,hirschmann15}; there is first rapid, early in-situ star formation creating building blocks that later merge together to assemble the galaxies we have today. Initially, negative metallicity gradients are expected from in-situ star formation as gas in the center experiences higher levels of star-formation and more enrichment. Later mergers would leave their impact on the metallicity gradient. 
\citet{kobayashi04} conducted smoothed-particle hydrodynamics (SPH) simulations of the formation of elliptical galaxies and showed that galaxies that had undergone major mergers with mass ratios less than 5:1 had significantly shallower gradients than galaxies that had undergone only minor mergers. 

The explanation is simply that major mergers more effectively disrupt the orbits of stars and mix them up from different regions of the galaxy, thereby flattening the original gradient. \citet{dimatteo09} found a similar result in binary merger simulations of two equal-sized ellipticals. For galaxies that initially had equal metallicity gradients, the merger remnant had roughly 0.6 times the initial gradient. \citet{hirschmann15} confirms these results for regions far from the galactic center. On the other hand, minor mergers could have the opposite effect and even supplement existing negative metallicity gradients since accreted stars would likely have low metallicities (coming from a low mass satellite galaxy) and tend to be added to the outlying regions of the host galaxy \citep{hirschmann15}. In the case of dissipational mergers, the effect on metallicity gradients is more complicated. On the one hand, metallicity gradients could be rejuvenated by a star-forming episode induced by the merger in the remnant's center \citep{navarro13,kobayashi04}. Specifically in modeling for shells galaxies, \citet{weil93}  found that in a dissipational minor merger, the gas and stars separate and the gas sinks to the center initiating a star-forming episode. On the other hand, gas that was originally far from the center of the galaxy could flow into the center during a major merger event \citep{rupke10}. Since gas far from the center would naturally be less metal enriched, this process can contribute to the metallicity gradient flattening and to a nuclear under-abundance of metals. Interacting galaxies have been shown to deviate to lower central metallicities from the mass-metallicity relation exhibited for normal galaxies \citep{ellison08}. Along these lines, \citet{longhetti00}  found that their sample of shell galaxies had smaller central Mg$_{2}$ index values than those predicted from the Mg$_{2}$-$\sigma$ relation from \citet{bender93}. 
	There have been several attempts at determining the ages of shell structures. \citet{canalizo07} use an N-body simulation of a minor merger to match the observed shells of quasar MC2 1635+119 estimating that a minor merger occurred around 1Gyr ago. This is connected to the presence of a roughly 1.5 Gyr stellar population in the center. \citet{hau99} date the shells of NGC 2865 by fitting measured index values with a `burst+bulge' model. They find that the measured index values are well fit by a roughly 0.4 Gyr population superimposed on an older bulge population. This young subpopulation is associated with the KDC in NGC 2865 and likely comes from the same wet merger event that created the shells. Determining the age of shells from stellar population studies could help refine and constrain the N-body simulations. 

This report is structured as follows: in Section~\ref{sec:datareduction} we describe the sample set and the data reduction, in Section~\ref{sec:methods}  we outline the methods for analyzing the stellar populations,  in Section~\ref{sec:results}  we discuss the results, and conclude in Section~\ref{sec:conclusions}.

\section{Data} \label{sec:datareduction} 

The data used for this project was taken in March 1996 with the Royal Greenwich Observatory spectrograph on the AAT. Details of the observation and the instrumental setup are listed in table 1. The sample of galaxies and some of their characteristics are given in table 2. 

The data reduction is fully explained in Hau et al. (in preparation). Exposures were typically 1800 seconds. Spectra were taken along the photometric minor and major axes of the galaxies. Some galaxies also had a spectrum taken along an intermediate axis. Rows along the slit were binned so that there was a minimum S/N of 30. 

\section{Methods}
\label{sec:methods} 
	To study the stellar populations of these galaxies, we implement two methods. We first analyzed the Lick indices and compared these to simple stellar population (SSP) predictions to derive SSP-equivalent ages and metallicities for the stars in the galaxies. SSP-equivalent parameters should roughly translate to luminosity-weighted average parameters. Second, we fit the whole galaxy spectra with high resolution SSP models. Since the full galaxy spectra have more information in them than individual absorption indices, we found the best fitting linear combination of SSP models. From this, we can directly infer the star formation history (SFH) of these galaxies. We can also derive mass weighted and luminosity weighted average age and metallicity estimates for the stars in these galaxies. 
	
\begin{table}
\caption{Instrumental Setup }
\begin{center}
\begin{tabular}{|c|c|c|}
\hline
Data & March 1996\\ 
Spectrograph & RGO\\
Camera & 25cm\\
Grating (lines/mm) & 1200V\\
Detector & TEK 1024 CCD\\
Gain (e-/ADU) & 2.7\\
Readout noise (e-/pix) & 4.8\\
Scale along slit (arcsec/pix) & 0.77\\
Dispersion (\AA/pix) & 0.79\\
Velocity per bin (kms$^{-1}$) & 42.7\\
Wavelength range (\AA) & 4850-5610\\
Instr FWHM (\AA) & 2.1\\
\hline
\end{tabular}
\end{center}
\label{tab:gals}
\end{table}
\begin{table}
\caption{Sample galaxies properties. A H$_0$ value of 70 \kms Mpc$^{-1}$ is adopted. Sources: 
1 - Hyperleda  }
\begin{center}
\begin{tabular}{|c|c|c|}
Name & $M_B$$^{1}$  \\
\hline
IC 2977 & -20.88 \\ 
IC 4329 &  -22.11\\
NGC 2945 & -21.07 \\
NGC 3051 & -20.20\\
NGC 3706 & -21.11\\
NGC 3923 & -21.25 \\
NGC 4936 & -21.62\\
NGC 5670 & -20.91\\
NGC 6776 & -21.70\\
\hline
\end{tabular}
\end{center}
\label{tab:gals2}
\end{table}%
\subsection{Emission Correction}
Before doing either the Lick analysis or the full-spectrum fitting, we cleaned our spectra of emission. Significant levels of emission are found within 6/9 of the galaxies (Hau et al. in preparation). We used the GANDALF algorithm \citep{sarzi06} in tandem with pPXF \citep{cappellari04} to simultaneously find the best combination of Gaussian emission lines and MILES SSP models \citep{vazdekis10} to fit our spectra. The best fitting emission lines were subtracted from each row in our spectra. Emission in H$\beta$, OIII, and NII were fit for and the H$\beta$ and NII lines were fixed to the kinematics of the OIII line. This was because OIII was generally the strongest line and therefore had the surest kinematics. All fits were visually inspected to guarantee the fit was reasonable. We investigated imperfect subtraction by measuring the Lick index values for the cleaned spectra and comparing them with the index values measured on the best fitting stellar template combination. Because of the large number of input SSP models, we would expect the stellar template to match the cleaned spectra fairly well and a large difference in the index measurements would indicate imperfect emission subtraction. For all of our galaxies except IC2977, the difference was within the noise level. For IC2977, emission was very strong and slightly non-Gaussian resulting in incomplete subtraction. To reflect this, we incorporated the difference in the index values between the cleaned spectra and the best stellar template as an estimate of the uncertainty in the measured index values due to the imperfect subtraction.  
\subsection{Lick Calibration}
Comparing index values with predictions from SSP models in the Lick system requires careful calibration. The calibration is three-fold. First, we degraded our spectra to the resolution of the Lick system. Even though the resolution of the spectra used in the Lick system is wavelength dependent \citep{worthey97}, the wavelength range of our data is small enough that a constant resolution can be assumed. We found the right amount to blur our spectra by blurring our observations of a standard star until they visually lined up with the observations of that star from the Lick library. Second, we had to correct for the different continuum shape due to the different flux calibration between our data and the data used in the Lick system. This was done by comparing index measurements on the one standard star in the Lick system that we have observations of (HD 54810) to the index measurements of that star in the Lick system. We derived additive offsets for each of the indices. Due to the fact that we only have one star in common with the Lick system, these corrective terms are a little uncertain. They were relatively small, however, and would not greatly affect the results. Additionally, the rest wavelength of the H$\beta$ index is on the edge of the spectral range of our data so we could not measure it for the standard star and hence could not derive a corrective offset (an offset of zero was assumed). Finally, we have to correct for the physical velocity dispersion in the galaxy spectra. To model how dispersion affects each of the indices, we artificially dispersed the spectra of 15 standard stars to varying levels of dispersion. From this, we could model how each of the index measurements changed as a function of dispersion. We derived a lookup table for each of the indices to be able to correct for any dispersion between 0 and 380 km/s. Again, the H$\beta$ index is just outside of our spectral range and so we took our dispersion correction for H$\beta$ from \citet{kuntschner98} who also used the RGO spectrograph (albeit with a different slit width). Clearly, the calibration is fairly uncertain but the results agree well with previous measurements for the galaxies that have been observed before (see $\S$4). 
There are 8 indices defined within our spectral range: $\hbeta$, Fe$_{5015}$, Mg$_{1}$, Mg$_{2}$, Mg$_{b}$, Fe$_{5270}$, Fe$_{5335}$, and Fe$_{5406}$. Index values were measured with the LECTOR program provided by A. Vazdekis \citep{vazdekis11}. 
Uncertainties in the index measurements were derived from photon statistics based on the formulae in \citet{cardiel98}. 
\subsection{Deriving SSP-equivalent Parameters}
To derive SSP-equivalent ages, metallicities, and alpha enhancements for our galaxies, we compared the index measurements to the SSP model predictions of \citet{thomas03}. These models extend in age from 0.1 to 15 Gyr, [Z/H] from -2.25 to 0.67, and [$\alpha$/Fe] from -0.3 to 0.5. The models were linearly interpolated to a regular grid with age spacing of 0.1 Gyr, metallicity spacing of 0.05 dex, and [$\alpha$/Fe] spacing of 0.05 dex. To find the best fitting age, metallicity, and alpha enhancement, we minimized the $\chi^{2}$ distance between the measured values of the 7 indices used (Mg$_{1}$ was discarded since it made many of the fits unstable) and the predicted values for the indices over a 3D grid of age, [Z/H], and [$\alpha$/Fe]. Indices that were over 3$\sigma$ away from the predictions at the best fitting point were discarded and the fit was redone. Always at least 5 indices were fit. Uncertainties in the SSP-equivalent properties were derived through Monte-Carlo simulation. Index values were resampled from a distribution with the measured value as the mean and the uncertainty from photon statistics as the standard deviation. These derived uncertainties include only uncertainty from random errors and not any systematic error arising from the model choice. Quoted uncertainties are the rms values from 50 repeated simulations. These uncertainties were compatible with uncertainty values derived by finding the parameter intervals that gave a $\Delta\chi^{2}$ of 3.53 \citep{press07}. The median contributions of each index to the overall minimum $\chi^{2}$ are generally around 1 or less indicating that the best fitting point was usually within 1$\sigma$ of all measured indices.

\subsection{Full Spectrum Fitting}
We performed full spectrum fitting with pPXF \citep{cappellari04} over a SSP basis from the P{\'E}GASE-HR models \citep{leborgne04}. This process finds the best fitting linear combination of SSP spectra from a 2D grid of age and [Fe/H]. We used SSP's corresponding to 40 ages logarithmically spaced between 100Myr and 14Gyr and 7 metallicities from Z=0.0001 to Z=0.1. The models were built using a Salpeter IMF. Metallicity was converted to [Fe/H] using [Fe/H] = 1.024log(Z)+1.739 from \citet{bertelli94} which assumes solar abundance patterns. The P{\'E}GASE-HR models have solar abundances which contrasts with the models used in the Lick indices analysis. In the future, it may be possible to do full spectrum fitting with SSP models for non-solar abundance patterns and currently those models are being developed \citep[e.g.][]{coelho07}. The galaxy spectra were first roughly flux calibrated based on observations of two stars in the MILES library \citep{falcon11}. A tenth order polynomial was still included in the fit to account for remaining flux calibration issues and for possible s-distortion. Including a polynomial has been shown not to affect the accuracy of the fitting process \citep{koleva08_val}. The absorption line kinematics were simultaneously fit along with the templates. The kinematics were parameterized as a fourth order Gauss-Hermite function.
	
	A solution from the fitting procedure consists of the linear combination of SSP models that best fit the galaxy spectrum. To suppress spurious solutions and reduce the degeneracy of the problem, a smoothness constraint was enforced on the solution via the regularization feature of pPXF. The solutions are smoothed at the cost of the minimum $\chi^{2}$ value. In other words, the function that is minimized is the combination of the $\chi^{2}$ metric and a measure of how ``un-smooth" the solution is which is derived from its second derivative. Regularization is a common method in recovering physically reasonable star formation histories from full spectrum fitting \citep[e.g.][]{ocvirk06}. Following other authors \citep{onodera12,mcdermid15} we used the maximum smoothness that still gave a reasonable quality of fit. Following the suggestions in \citet{press07}, this was quantified as the fit that resulted in a minimum $\chi^{2}$ value of ~ N+(2N)$^{1/2}$ where N is the number of pixels fit. A minimum $\chi^{2}$ value of ~ N signifies a good fit and, as was stated before, the quality of the fit is sacrificed in favor of a smoother solution. But to reduce the risk of systematic error affecting derived age and metallicity gradients, the same level of regularization was used for each row in a galaxy spectrum. The regularization was chosen so that it was the maximum that still gave a good fit for all rows in the spectrum. 
	
\begin{figure*}
\includegraphics[height=0.45\textheight,width=0.9\textwidth]{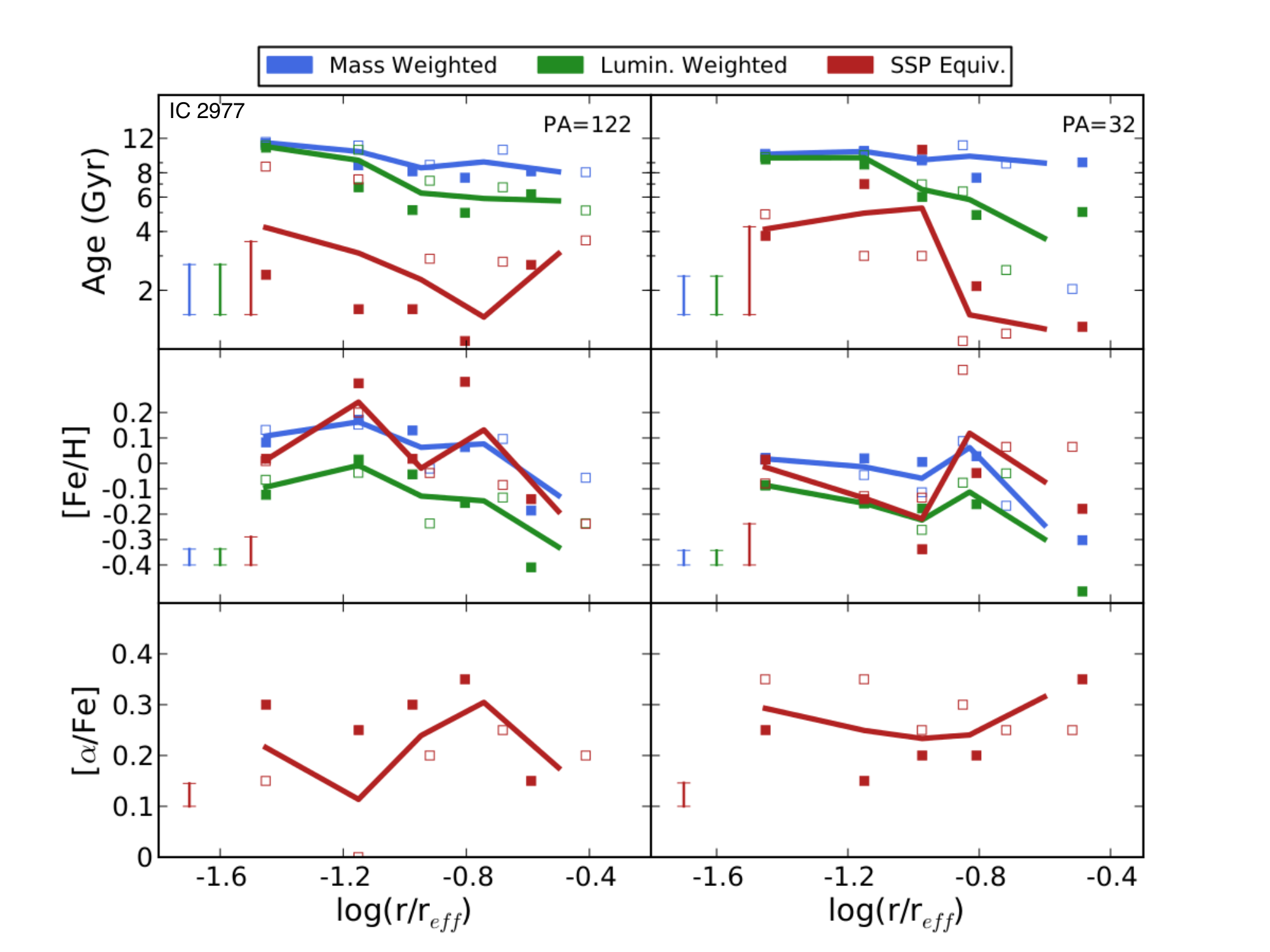}
\end{figure*}
\begin{figure*}
\includegraphics[height=0.45\textheight,width=0.9\textwidth]{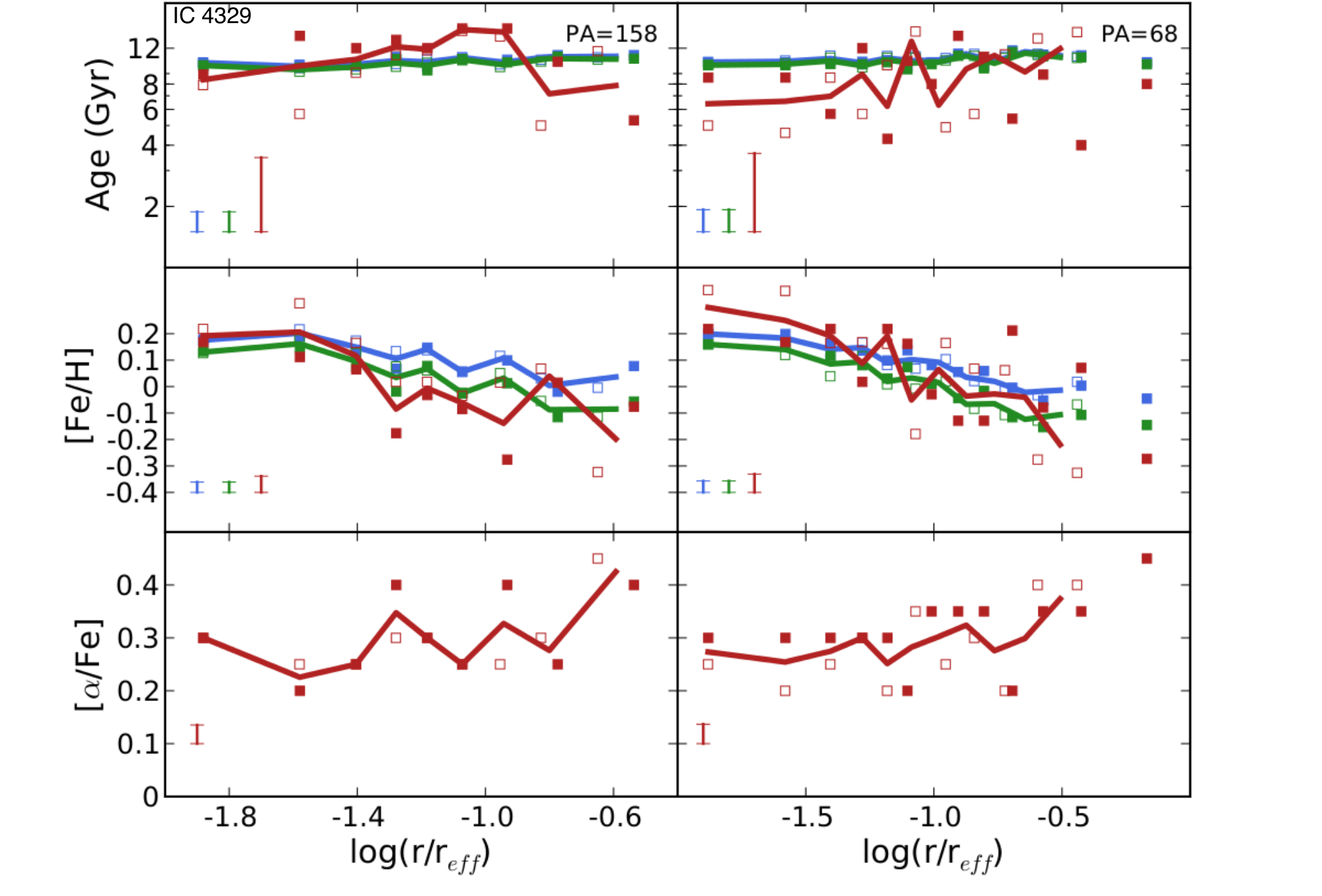}
\end{figure*}
\begin{figure*}
\includegraphics[height=0.45\textheight,width=0.9\textwidth]{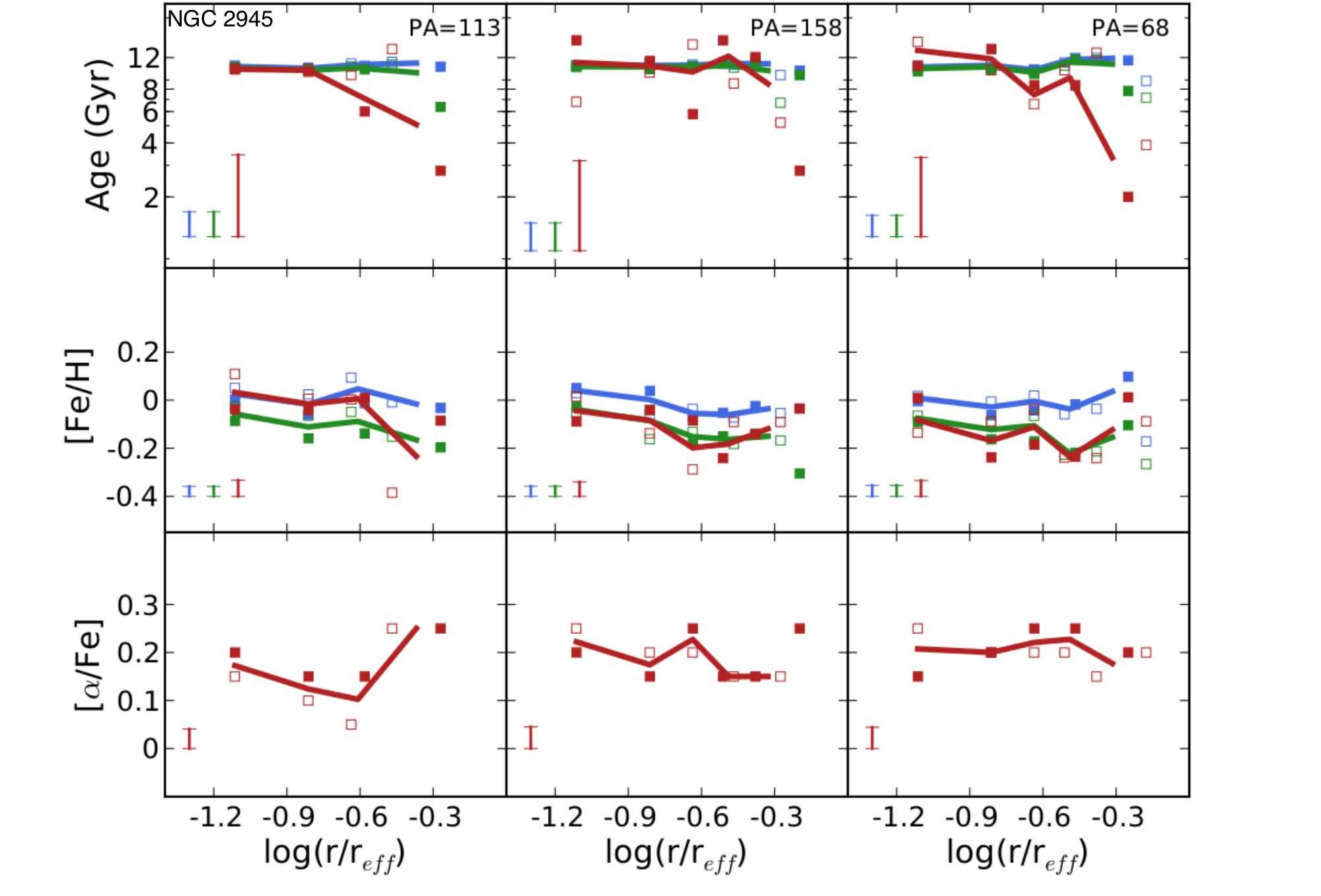}
\end{figure*}
\begin{figure*}
\includegraphics[height=0.45\textheight,width=0.9\textwidth]{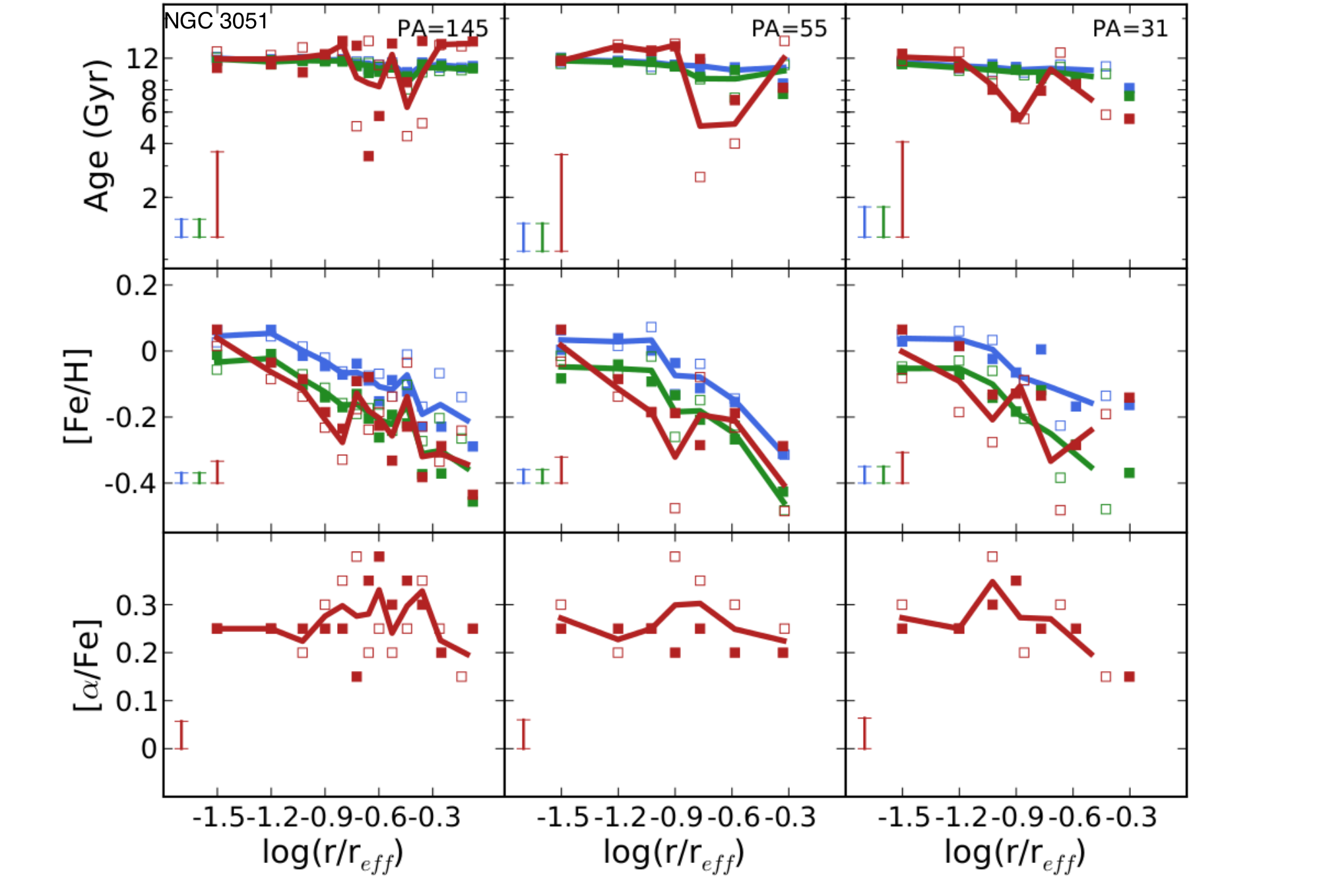}
\end{figure*}
\begin{figure*}
\includegraphics[height=0.45\textheight,width=0.9\textwidth]{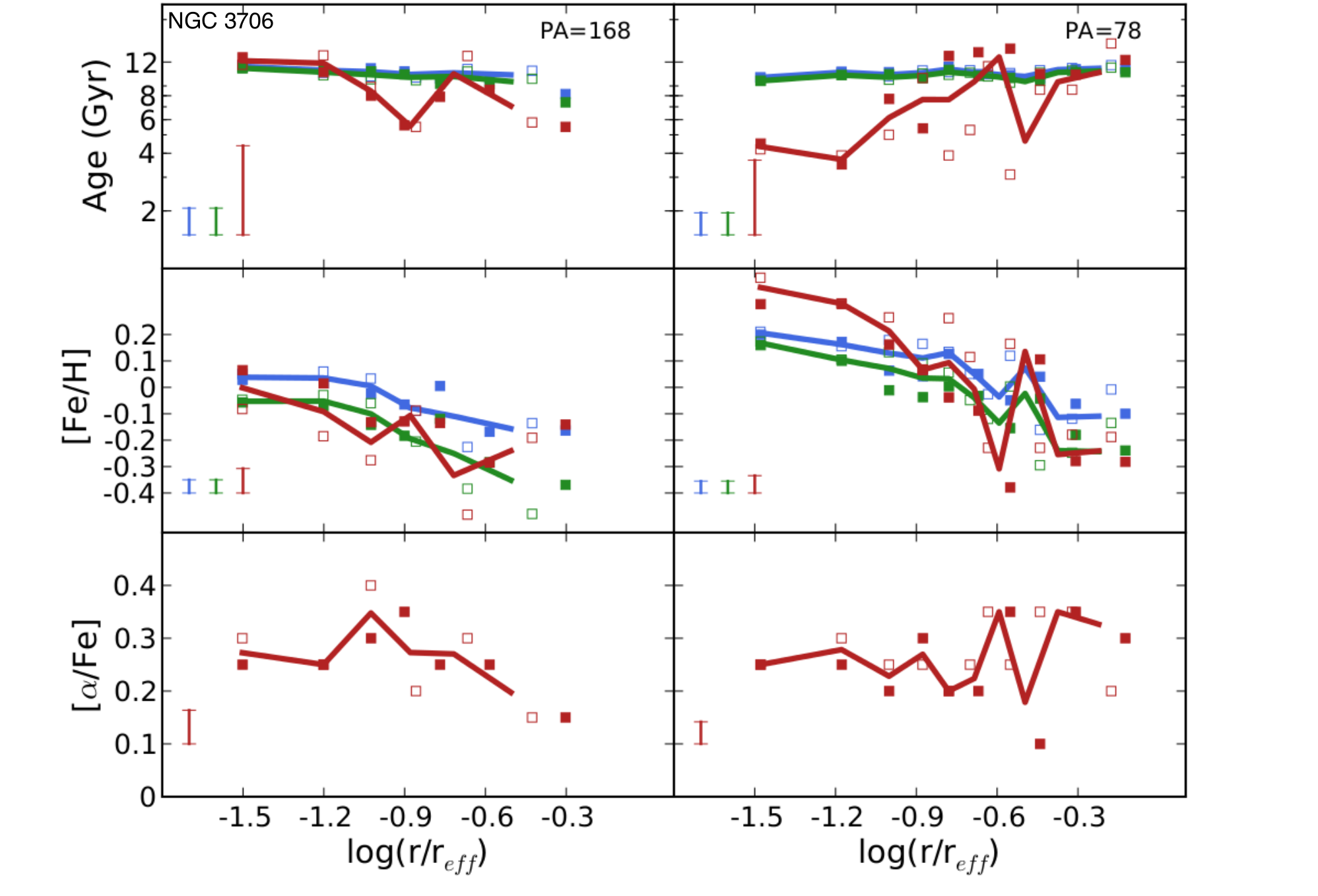}
\end{figure*}
\begin{figure*}

\includegraphics[height=0.45\textheight,width=0.9\textwidth]{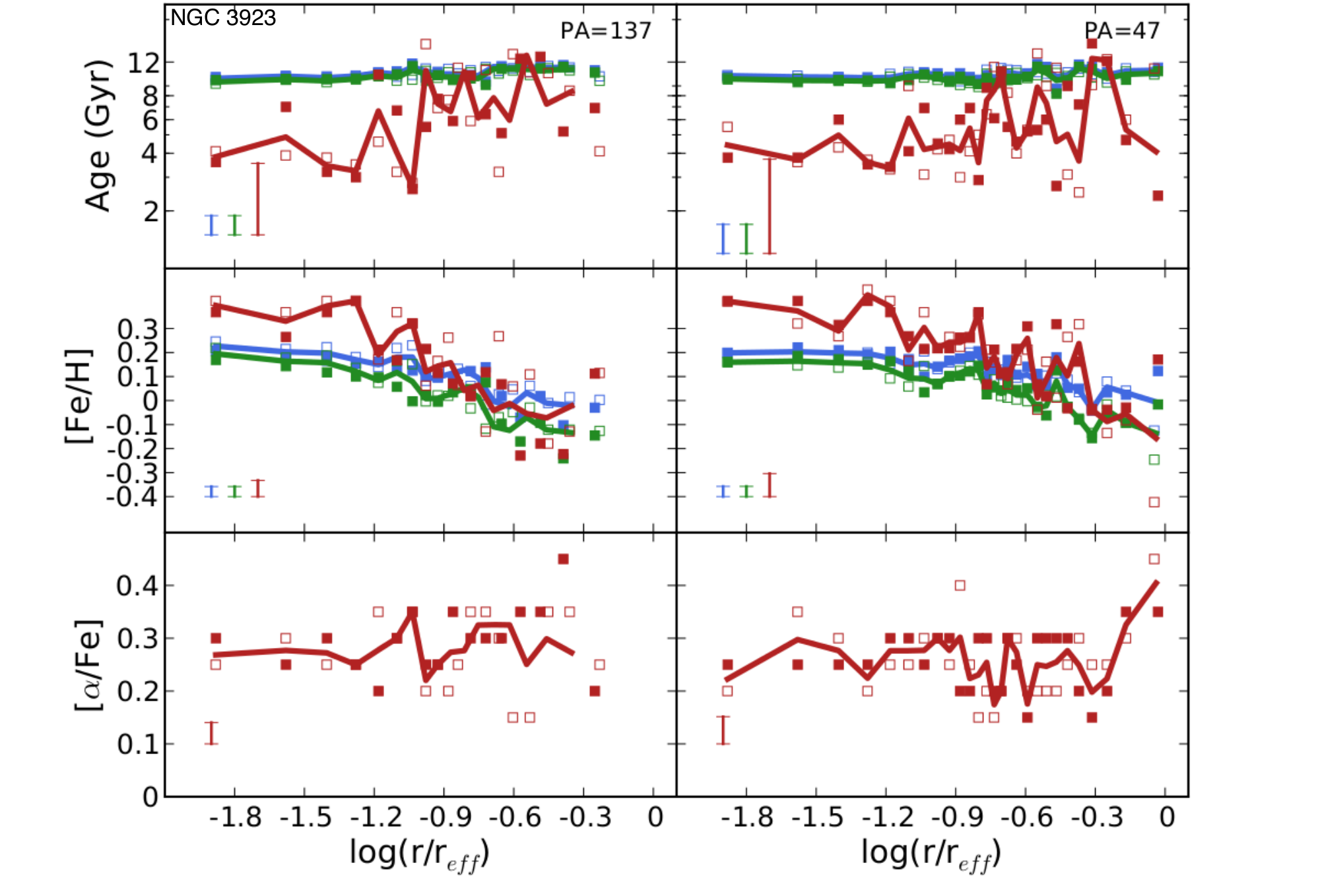}
\end{figure*}
\begin{figure*}
\includegraphics[height=0.45\textheight,width=0.9\textwidth]{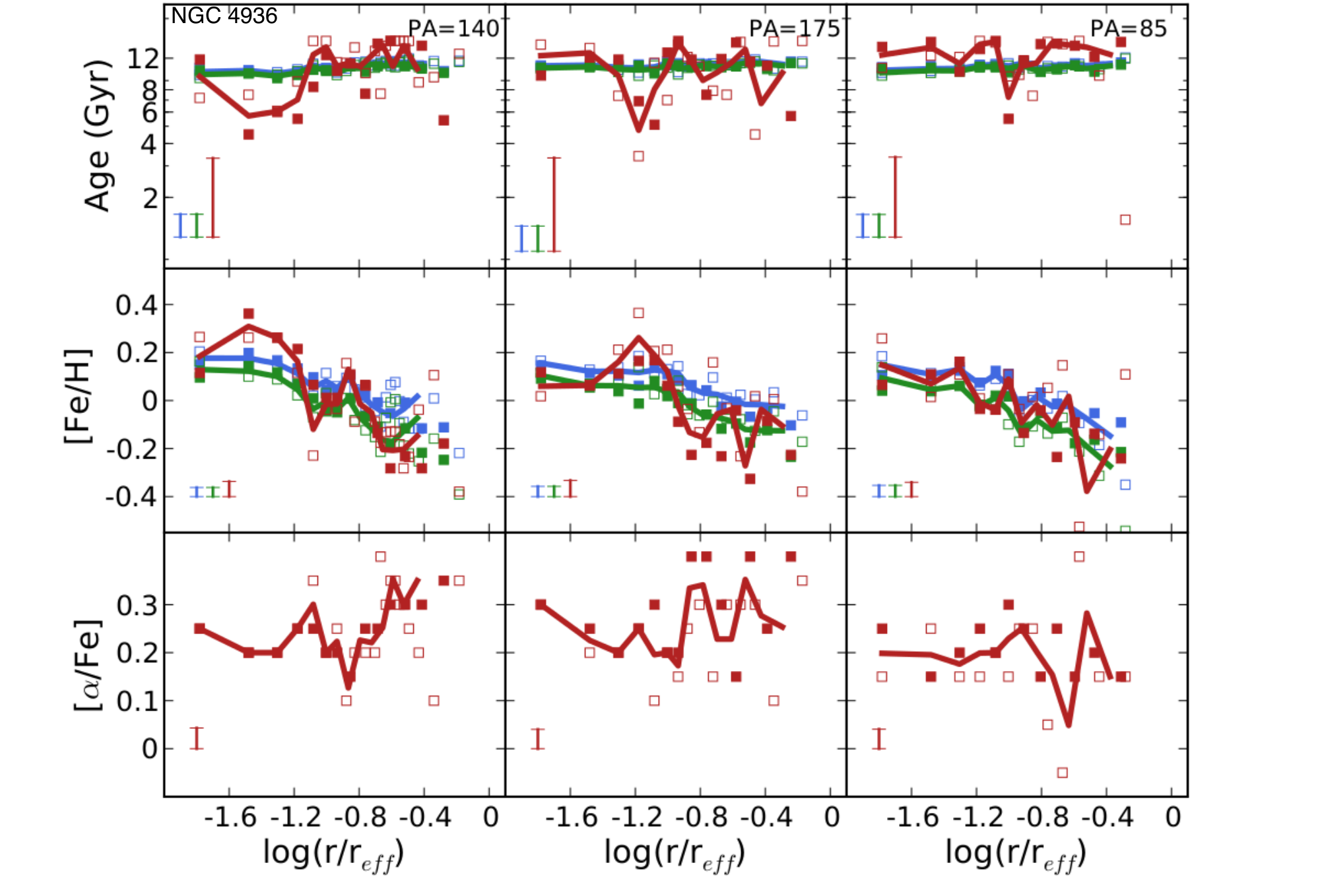}
\end{figure*}
\begin{figure*}

\includegraphics[height=0.45\textheight,width=0.9\textwidth]{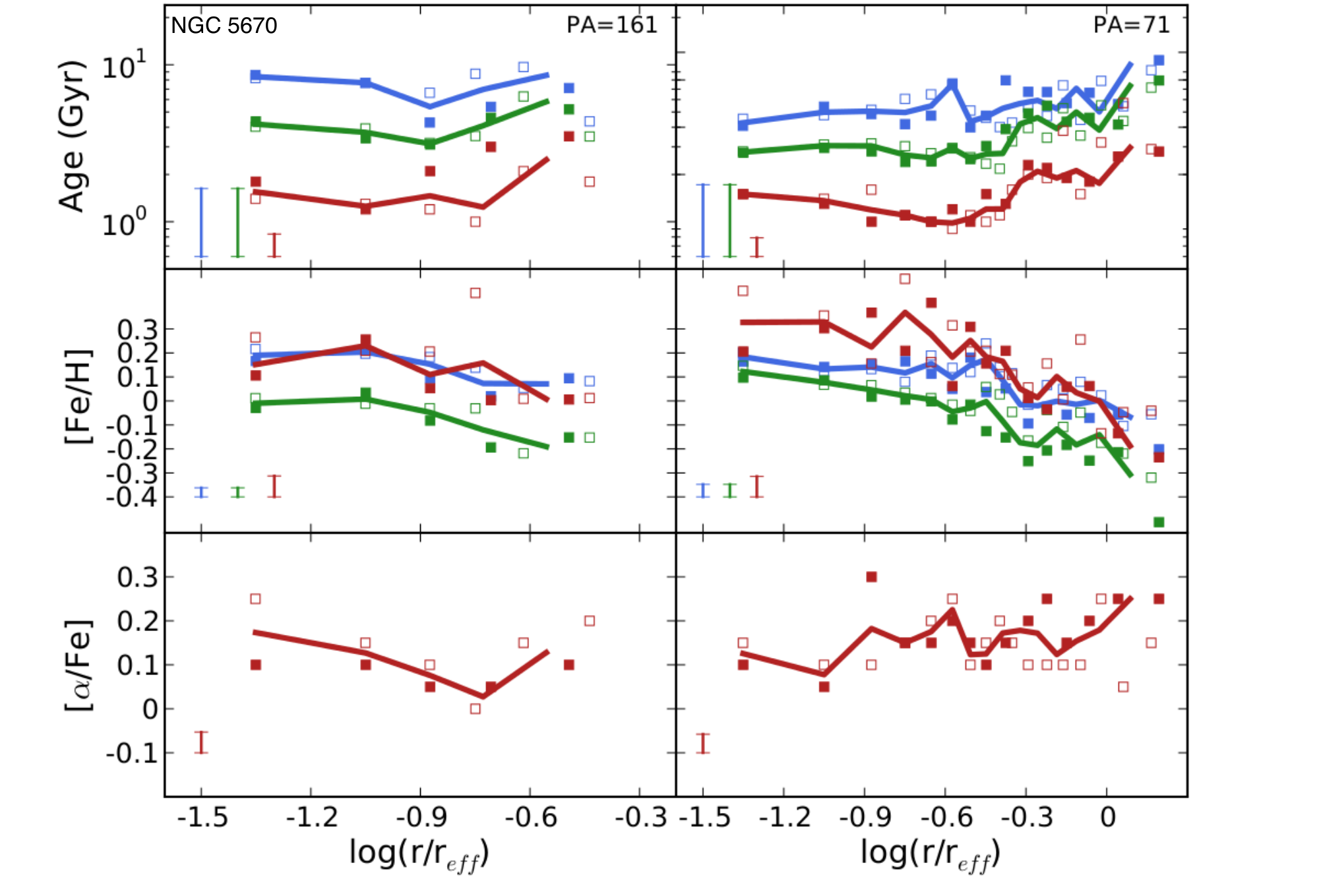}
\end{figure*}
\begin{figure*}
\includegraphics[height=0.45\textheight,width=0.9\textwidth]{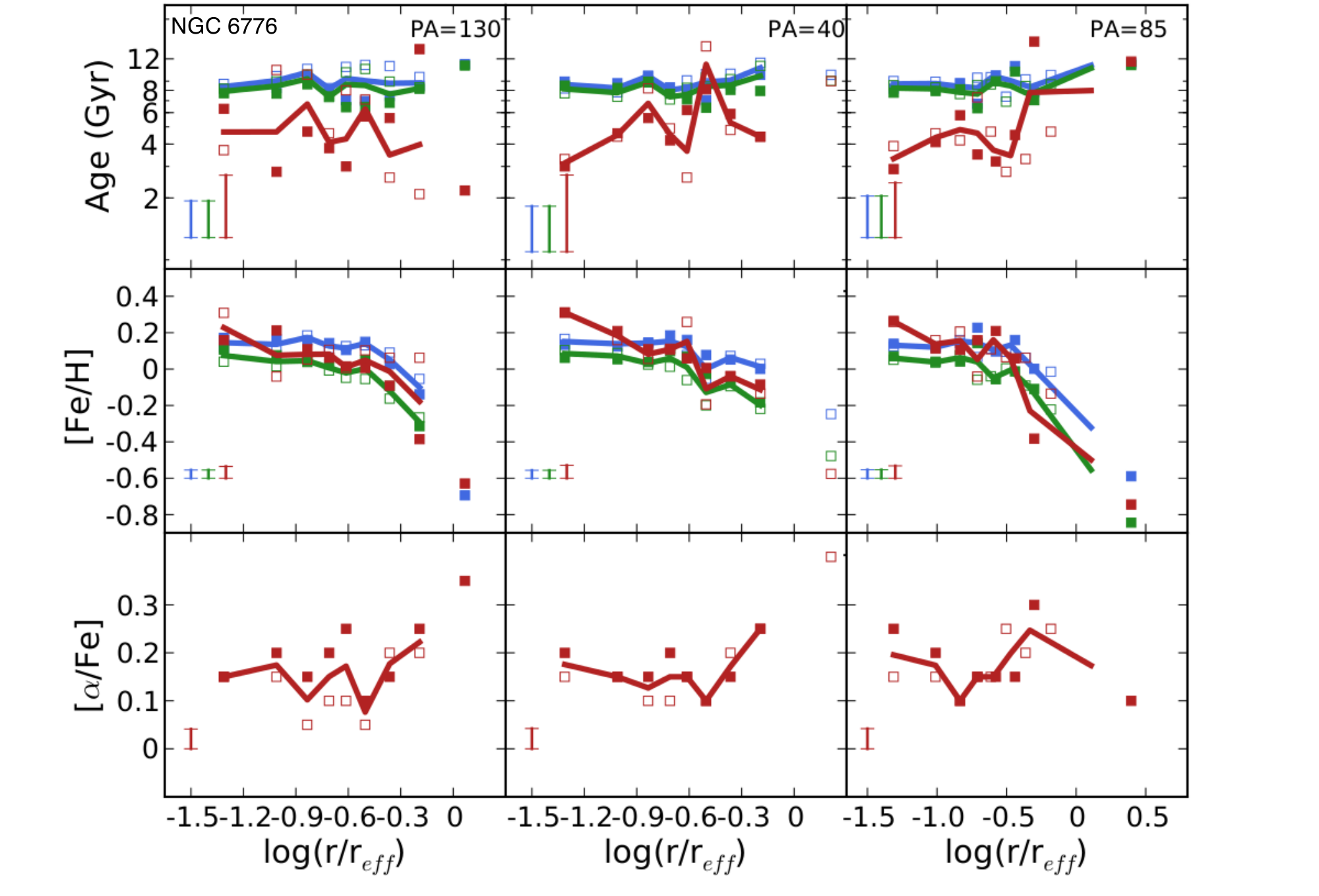}
\end{figure*}
\begin{figure*}
\caption{The  stellar parameters for the 9 shell galaxies. Values from the positive side of the slit axis are plotted as filled squares where those from the negative side are open squares. Blue points show the mass-weighted full-spectrum fitting results, green show the luminosity-weighted full-spectrum fitting results, and red show the SSP equivalent results derived from the Lick indices. The colored lines represent weighted averages of the data. The error bars represent the median uncertainty of the the population parameter for each of the three methods.}
\label{fig:popparams}
\end{figure*}

	The SSP models are normalized in luminosity to 1 solar mass of initial population size and thus the solution yields mass weights for each SSP. From the best fitting weights for a 2D grid of metallicity and age, we derived mass-weighted average age and metallicity. From the relative amplitudes of the SSP template spectra (which reflects the mass/light ratio of a population with a certain age and metallicity), we can also derive the luminosity weighted average age and metallicity. These averages were calculated as:
\begin{equation}
\langle \log(Age)\rangle = \frac{\Sigma w_{i} \log(Age_{i})}{\Sigma w_{i}}
\end{equation}
\begin{equation}
\langle [Fe/H] \rangle = \frac{\Sigma w_{i} [Fe/H]_{i}}{\Sigma w_{i}}
\end{equation}
where $w_{i}$ is the mass fraction or the luminosity fraction of a certain age and metallicity bin. We averaged the base 10 logarithm of the age because it tended to be more affected by young components which made the results more compatible with the index derived results. If we instead used an average of linear age, the average would be different but the difference is relatively small (typically ~1-2 Gyr). 
Uncertainties were derived through Monte-Carlo simulation whereby each pixel of each spectra was resampled from a Gaussian distribution of width equal to the observational error of that pixel. The resampled spectra were analyzed in pPXF with the same regularization as the original spectra. Average age and metallicity were calculated for the resampled spectra and quoted uncertainties are the rms values derived from repeating this procedure 50 times. 

\section{Results and Discussion}
\label{sec:results} 

The stellar population parameters derived from both methods are shown
in figure~\ref{fig:popparams}. Plotted together are the mass-weighted,
luminosity-weighted, and SSP-equivalent ages and $[Fe/H]$, along with the
SSP-equivalent $[\alpha/Fe]$. $[Fe/H]$ for the SSP-equivalent results is
derived from $[Z/H]$ using equation (4) of \citet{thomas03}:

\begin{equation} 
[Z/H] = [Fe/H] + A[\alpha/Fe]
\end{equation}
where, for the mixtures in the Thomas et al. (2003) models, $A=0.94$. 

As expected, the luminosity-weighted values from full-spectrum fitting
agree more closely with the SSP-equivalent results, but there are some
differences especially in age. The difference seems to be most
significant when the index fitting yields young ages. This is similar
to what \citet{sanchez11} found from studying the stellar populations
of bulges. There is clearly much more scatter in the index-derived
results, and the age/metallicity degeneracy can be seen. For example,
outliers which are high in metallicity have low derived ages, and {\it
  vice versa}. The large scatter is likely due to the relatively few
(7) indices that were used in the fit. Despite the large scatter, the
index-fitting results agree well with previous results. Our $\mathrm{Mg}_{2}$,
$\mathrm{Mg}_{b}$, $\mathrm{Fe}_{5270}$, and $\mathrm{Fe}_{5335}$ index values over the central
$5\arcsec$ agree well with those measured by \citet{longhetti00} for
the three galaxies in both samples: NGC 2945, NGC 3051, and NGC
6776. Our derived H$\beta$ index values did not agree well with those
of \citet{longhetti00}, but that is likely because their spectra were not
cleaned of emission and all three galaxies were found to have ionized
gas emission. The young central SSP-equivalent ages, high metallicity,
and high $\alpha$-enhancement that we find in NGC 3923 agree well with
previous results from \citet{thomas05} and
\citet{denicolo05}. \citet{denicolo05} find a lower
$\alpha$-enhancement, however. Both of these authors used the same
\citet{thomas03} models as we did, but they used a different method
for determining SSP-equivalent parameters involving index-index
planes.

It has been shown by several studies that full-spectrum fitting is
more accurate and precise at determining age and metallicity
\citep{koleva08_val,sanchez11}. It is also less susceptible to the
presence of young stellar subpopulations, which we find evidence for
in several of our galaxies. For these reasons, we adopt the ages and
metallicities given by full-spectrum fitting when discussing central
populations and population gradients. However, we use the
$\alpha$-enhancement to convert $[Fe/H]$ to total metallicity
$[Z/H]$. Since the $[\alpha/Fe]$ values are more stable than age or
metallicity and do not show steep gradients, they would not
affect the calculated metallicity gradients.

\subsection{Central Stellar Populations}

\begin{figure}
\includegraphics[scale=0.31]{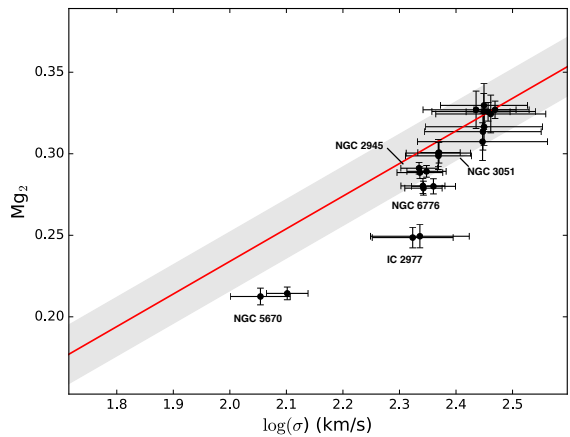}
\caption{Central Mg$_{2}$ vs $\sigma$. The red line shows the \citet{bender93} Mg$_{2}$-$\sigma$ relationship for normal early type galaxies. The gray region describes the intrinsic scatter that \citet{bender93} found for their relation. The five galaxies with the lowest velocity dispersion are labelled.}
\label{fig:centralMetallicity}
\end{figure}

Signatures of a galaxy merger can be found in the central part of galaxies.  It has been shown that normal galaxies exhibit tight relations between central velocity dispersion and certain line indices, particularly Mg$_{2}$ \citep[e.g.][]{bender93,graves09,greene12}. Deviations from these trends have been taken as indicators of galaxy interaction \citep{longhetti00}.
Figure~\ref{fig:centralMetallicity} shows a plot of the central Mg$_{2}$ index values from an R$_{\mathrm{eff}}$/8 extraction vs the central velocity dispersion for all position angles of our galaxy sample. They are plotted with the Mg$_{2}$-$\sigma$ relationship found by \citet{bender93} for normal early type galaxies. Our data show a clear trend whereby the Mg$_{2}$ index increases for increasing kinematic dispersion but appears to be slightly offset below the \citet{bender93} model. This result was also found by \citet{longhetti00} in their sample of shell galaxies. This effect is likely due to the merger history of shell galaxies. Previous mergers could disrupt the galaxies' kinematics and lead to increased velocity dispersion which would cause it to deviate from the regular Mg$_{2}$-$\sigma$ relation. Previous mergers could also increase star formation and bring the overall Mg$_{2}$ index value down due to the presence of bright, young stars. In fact, the very low dispersion galaxy shown in figure $\ref{fig:centralMetallicity}$ which seems to deviate the most from the model relation is NGC 5670 which has the youngest ages from the Lick indices and full-spectrum fitting. IC 2977 also significantly deviates from the relation shown in Figure~\ref{fig:centralMetallicity}, however, this is likely due to an inaccurate Mg$_{2}$ value due to incomplete subtraction of the gas emission, as described earlier.

\begin{figure}
\includegraphics[scale=0.45]{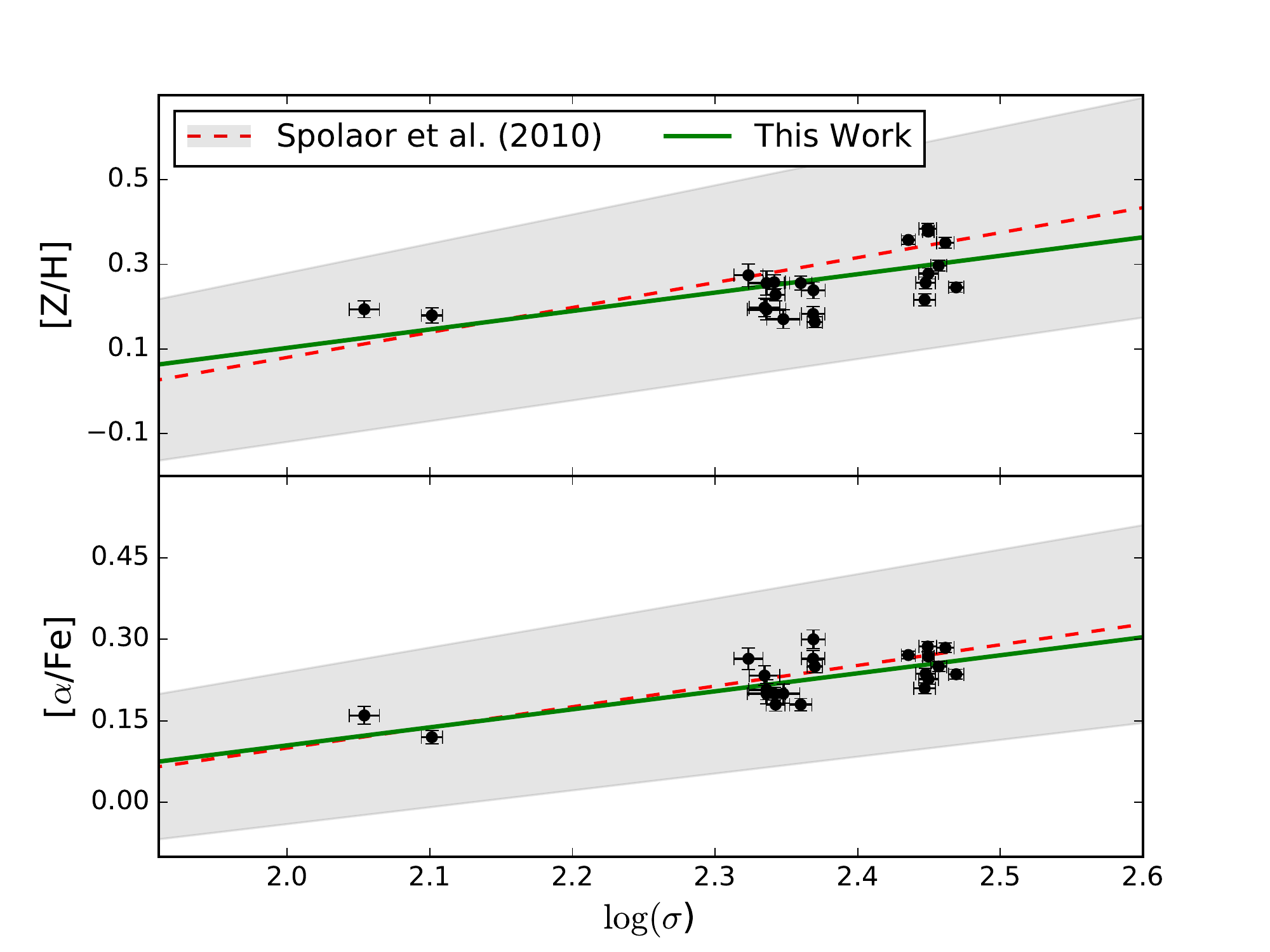}
\caption{$[Z/H]$ and $[\alpha/Fe]$ plotted against dispersion $\sigma$. All values are averaged over the central $R_{\mathrm{eff}}/8$ portion of each galaxy.}
\label{fig:spolaor}
\end{figure}

Figure~\ref{fig:spolaor} shows a plot of the central $[Z/H]$ and $[\alpha/Fe]$ as a function of velocity dispersion for all position angles of our galaxies. All values are averaged over a central $R_{\mathrm{eff}}$/8 aperture. The $[\alpha/Fe]$ is derived from the Lick indices whereas the $[Z/H]$ is derived from the central mass-weighted $[Fe/H]$ from full-spectrum fitting and the central $[\alpha/Fe]$ using equation (3) above. $[Z/H]$ is derived this way and not directly from the SSP-equivalent $[Z/H]$ because the full-spectrum fitting results are more accurate and precise for reasons explained before. Also shown in figure~\ref{fig:spolaor} are the $[Z/H]$ and $[\alpha/Fe]$-$\sigma$ relations found by \citet{spolaor10} 
along with the 1$\sigma$ confidence region for their relation. The \citet{spolaor10} sample includes galaxies that span a wide range in masses and environments and can be considered as `normal' early type galaxies. Our [$\alpha$/Fe]-$\sigma$ relation appears to line up well with the \citet{spolaor10} relation. While within the uncertainty of the \citet{spolaor10} relation, our galaxies do seem to be displaced $\sim$ 0.05-0.1 dex below their $[Z/H]$-$\sigma$ relation. A displacement from the normal mass-metallicity relation to lower metallicity has been found for interacting galaxy pairs \citep{kewley06,ellison08}. This shift has been explained as low metallicity gas from the outskirts of the galaxy being funneled into the center \citep{rupke10,kewley06}.
The deviation of our sample from the \citet{spolaor10} relation could also be due to an increased velocity dispersion resulting from mergers. 


\subsection{Population Gradients}
\begin{figure*}
 \includegraphics[scale=0.62]{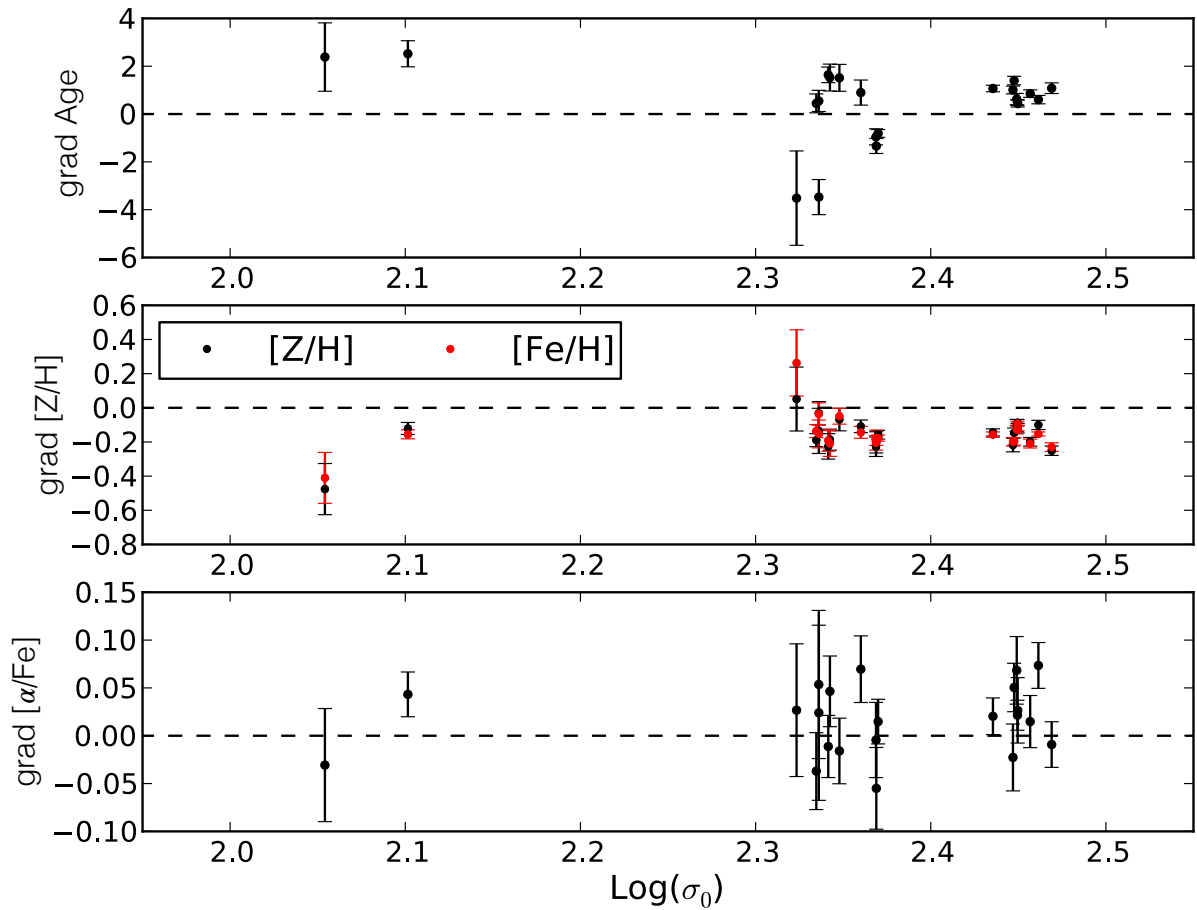}
\caption{Stellar population gradients plotted against the central velocity dispersion $\sigma_0$, averaged over the central $R_{\mathrm{eff}}/8$.}
\label{fig:popgradient}
\end{figure*}

To study the stellar population gradients, We performed linear fits to
the parameters of the form:
\begin{equation}
{\rm P}(r) = {\rm P}_{r_{\mathrm{eff}}} + \frac{\Delta {\rm P}}{\Delta {\rm log}\ r} {\rm log} \frac{r}{r_{\mathrm{eff}}}
\end{equation}
where $r$ is the galactocentric radius, $r_{\mathrm{eff}}$ the effective radius, P$_{r_{\mathrm{eff}}}$ is the stellar population parameter evaluated at $r_{\mathrm{eff}}$, and $P$ is any of the three measured parameters: metallicity, age,
or $\alpha$-element enhancement.  The gradients for age and metallicity came from the mass-weighted full-spectrum fitting, while
those for the $\alpha$-element enhancement gradient came from the
Lick indices analysis. The metallicity gradients are calculated both
as $[Fe/H]$ and $[Z/H]$. As discussed above, the $[Z/H]$ calculation
uses the SSP-equivalent $\alpha$-enhancement results.  The fits were
weighted according to the uncertainties in each of the data points.
Figure~\ref{fig:popgradient} shows the derived gradients plotted against the velocity
dispersion $\sigma_0$ (averaged over the central $R_{\mathrm{eff}}/8$).

The age and $\alpha$-enhancement gradients are generally small and
around zero, while the metallicity gradients are negative, as
expected. Except for one position angle of IC 2977, there is little
difference between gradients in $[Fe/H]$ and $[Z/H]$. There does not
appear to be any correlation between $\sigma_0$ and any of the three
gradients. This is consistent with previous studies
\citep{mehlert03,koleva11}. With that said, it is interesting to note
that the steepest metallicity gradient comes from the galaxy with the
smallest dispersion in our sample, NGC 5670. Other authors have found
a complicated relation where the gradient is steepest around
$\log(\sigma) \sim 2.1$ \citep{sanchez07,spolaor10}, which
corresponds roughly to the central dispersion of NGC 5670.

As discussed above, metallicity gradients give clues to the formation
and merging history of galaxies. Gradients exist because gas in the
center of a galaxy is enriched due to more efficient star formation,
compared to gas further out.  Gas that is more tightly held undergoes
more star formation before supernovae and galactic winds drive it out.

Our sample of galaxies have an average metallicity gradient along
their major axes of $\Delta [Z/H]/\Delta \log(r) = -0.16 \pm 0.10$ dex
decade$^{-1}$ in radius. This is comparable to, but less than, those
found in similar studies \citep{sanchez07,koleva11} of regular galaxies.  Gradients of
$\sim -0.30$ to $-0.50$ dex decade$^{-1}$ are expected from monolithic
collapse models \citep{pipino10,pipino08}, significantly steeper than
those we find. To explain the shallow gradients we find, galaxy mergers are needed to mix up the stars and
dilute the gradient. 

In  SPH simulations which included many feedback processes including supernovae and
star formation, \citet{kobayashi04} found that galaxies that had
undergone major mergers have significantly shallower gradients than
those that had only undergone minor mergers (mass ratio $>$ 5:1). The
major and minor merger groups have average metallicity gradients of
$\Delta[Z/H]/\Delta \log(r) = -0.22$ and $-0.30$ respectively. The \citet{kobayashi04} results are V-band luminosity weighted
gradients. Fitting a gradient to our luminosity 
weighted results yields a slightly steeper average
metallicity gradient of $\Delta [Z/H]/\Delta \log(r) = -0.22 \pm 0.11$ dex decade$^{-1}$. This is still suggestive of major mergers but is also consistent with minor mergers.
A possible issue with comparing with
the Kobayashi (2004) results is that the models used in those
simulations are based on elemental abundances in the solar
neighborhood. Abundances are fixed for a certain $[Fe/H]$: for $[Fe/H]
\leq -1$, the abundances of galactic halo stars are used
($[O/Fe]=0.5$); for $[Fe/H] \geq 0$, solar abundances are used; and
for intermediate $[Fe/H]$ values, abundances were interpolated from
the two extremes. In our results, we find the presence of metal rich
stars with $[Fe/H] > 0$ that are $\alpha$-element enhanced. We find an
average [Fe/H] gradient $\Delta [Fe/H]/\Delta \log(r) = -0.15 \pm 0.11$
dex decade$^{-1}$, much shallower than the iron gradients of $-0.45$
and $-0.38$ that \citet{kobayashi04} found for the two groups --
non-major mergers and major mergers. In summary, our galaxies do
show very shallow metallicity gradients and many have likely been in a
major merging episode in the past.  It is difficult to deduce,
however, whether shells formed during these episodes or during
unrelated, minor mergers.

It is worth noting that NGC 3923 was considered a good candidate for
having its shells formed via a major merger by \citet{hernquist92} due
to the large range in shell radii. The very shallow metallicity
gradient of $\Delta [Z/H]/\Delta \log(r) = -0.08 \pm 0.018$ dex
decade$^{-1}$ that we find would seem to confirm this.  However
\citet{norris08} derived SSP-equivalent stellar population parameters
for a point at $\sim 3r_{\mathrm{eff}}$, and found an old age of $\sim 15$ Gyr
and a poor metallicity of $[Z/H] = -0.33$ and $[Fe/H] = -0.65$. This
would imply that the metallicity gradient steepens significantly
between $r= r_{\mathrm{eff}}$ and $3r_{\mathrm{eff}}$. We can see some steepening happening in our data outwards of $\sim$0.3r$_{\mathrm{eff}}$. Within this radius, it appears that the metallicity plateaus. This flattening has been found in other massive galaxies,
such as M87 \citep{montes14} and NGC 1600 \citep{sanchez07}. The
former consider the plateau to be due to a very short and intense
burst of star formation that formed the core in early times, while
later accretion of other galaxies built up the rest of the
galaxy. This scenario could explain the gradients in NGC 3923 if the
accreted stars had low metallicities and were added to the outskirts
of NGC 3923 \citep{hirschmann15}, giving rise to a steepening of the
metallicity gradient. The shells could have formed during one of these
accretion events. To confirm that the metallicity gradient does get
steeper further out, more observations of outer regions are necessary.
	
\begin{figure*}
 \includegraphics[scale=0.48]{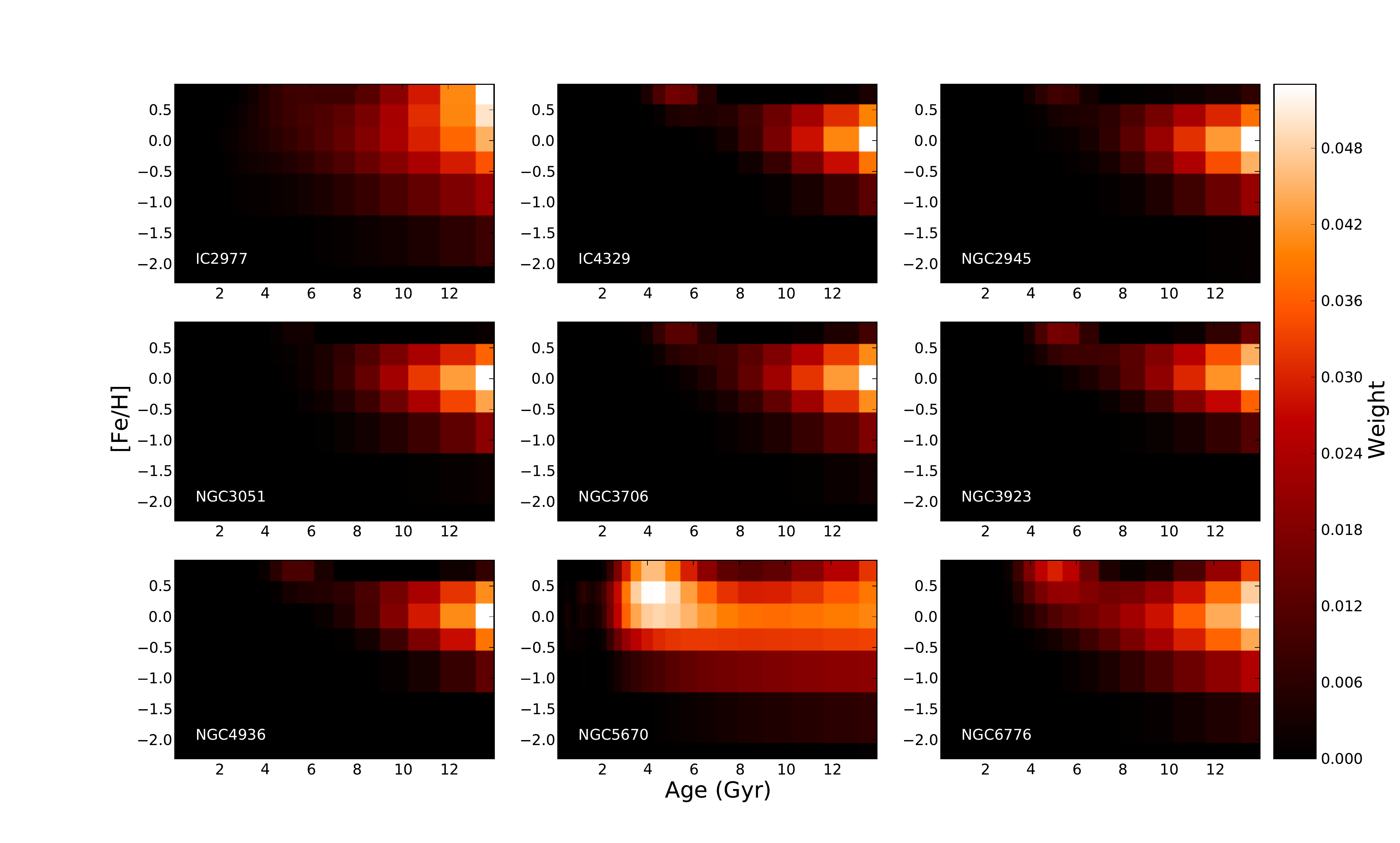}
\caption{Mass weight distribution in $[Fe/H]$ vs ($Age$) for all nine galaxies.}
\label{fig:massweight}
\end{figure*}

\subsection{Young Stellar Populations}

\begin{figure}
 \includegraphics[scale=0.2]{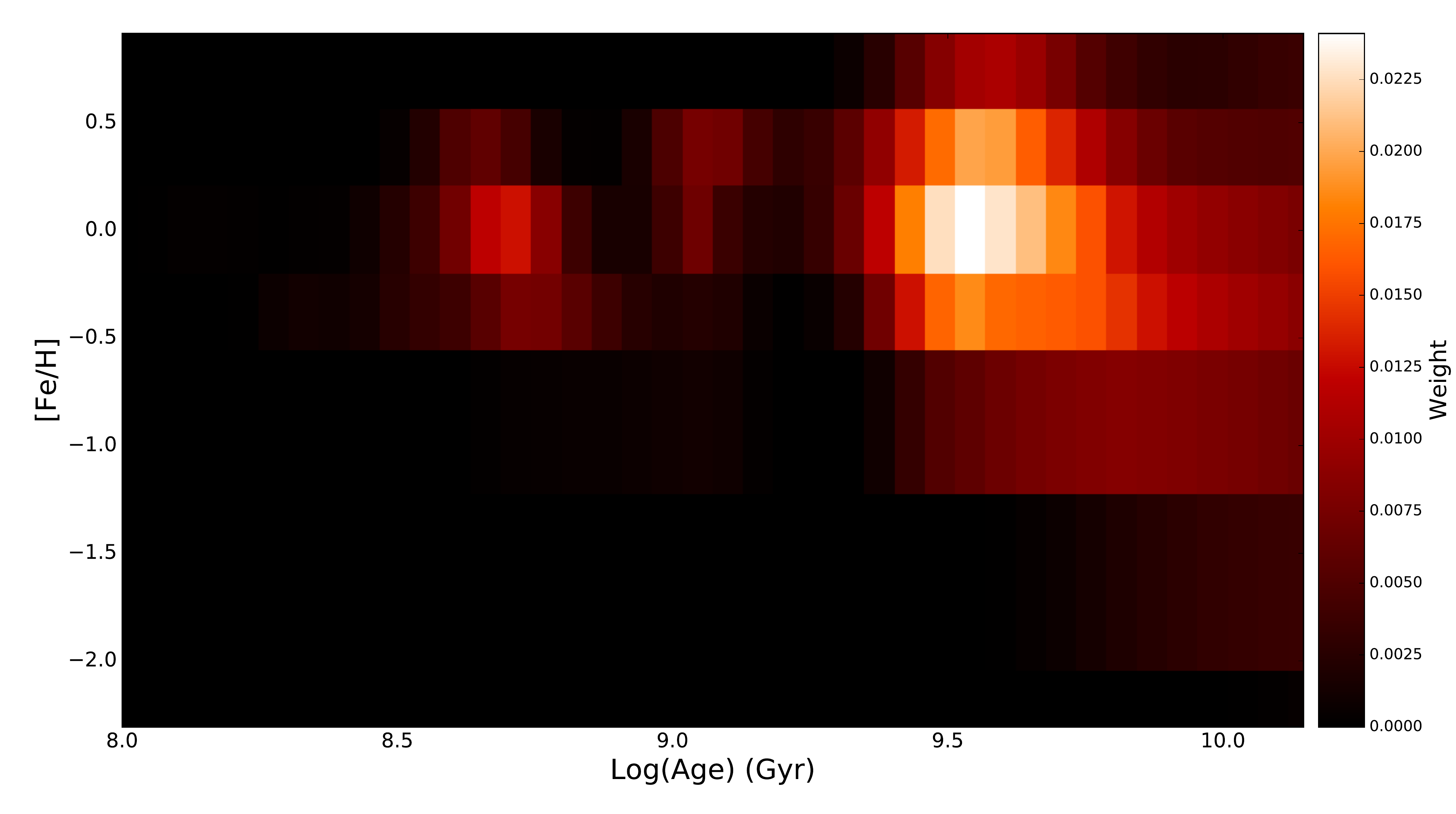}
\caption{Weight distribution for NGC 5670 this time shown with log(Age) on the horizontal axis and now showing the luminosity weights for each SSP instead of the mass weights. A bright, young subpopulation of age $\sim$ 500 Myr is clearly visible.}
\label{fig:massweight5670}
\end{figure}

We find evidence for young subpopulations in several of the
galaxies. Figure~\ref{fig:massweight} shows the weights found from full-spectrum fitting
for the different age and metallicity bins for each galaxy's major
axis averaged over the central $R_{\mathrm{eff}}$/8. With the exception of NGC
5670, most galaxies are dominated by old stars ($\sim$ 10 Gyr old) with roughly solar
metallicity. 
NGC 5670, the least massive galaxy in our sample, appears
to be dominated by a roughly 5-6Gyr old population. 
Figure~\ref{fig:massweight5670} shows a
plot of the assigned weights for just NGC 5670. Plotted against
log(Age), a very young ($\sim$400-500Myr) population is visible.  
Four of the other galaxies appear to have young ages from the Lick
analysis. These are NGC 3923, NGC 3706, NGC 6776, and IC 2977 
with central ages around 4 Gyr. As was stated in $\S$3.1, IC 2977 had the
strongest gas emission and its index measurements are likely affected
by imperfect subtraction of the emission lines. Therefore, the Lick
results for IC 2977 are likely skewed. The other three galaxies: NGC
3923, NGC 3706, and NGC 6776 show a ~4Gyr old, very metal-rich
subpopulation in the full-spectrum results as well. These
subpopulations are seen in all Monte Carlo realizations of the data
and so are unlikely to be due only to random noise in the data. They
are also present over a relatively wide range of values for the
regularization parameter. However, too low or too high of a
regularization parameter and they are not seen. This is to be expected
because at too high of regularization details in the SFH are blurred
out and at too low of regularization spurious solutions dominate. The
young ages and high metallicities of these subpopulations are in good
agreement with the ages and metallicities inferred from the Lick
indices. \citet{serra07} found that, for a composite stellar
population, the SSP-equivalent ages tend towards the age of the young
subpopulation whereas the SSP-equivalent metallicity tended towards
the metallicity of the dominant older population. In our case,
however, the SSP-equivalent ages $\textit{and}$ metallicities seem to
tend to those of the young subpopulation. We conducted a simple test
by constructing a composite stellar population of one old,
solar-metallicity spectrum from the P{\'E}GASE-HR models and one
young, metal-rich spectrum in roughly the proportion found for NGC
3923 (with a $\sim$5:1 ratio between the old and young components). We
artificially dispersed this model to match the velocity dispersion of
NGC 3923 and then ran it through the Lick analysis. The results were
reasonably consistent with the SSP-equivalent results we actually
found for NGC 3923 except that the model had no $\alpha$
enrichment. This test was simplistic but is suggestive that the young
subpopulations could bring about a match between the index-derived
results and the full-spectrum fitting results. With that said, the
fact that a similar $\sim$4-5 Gyr population was found in so many of
the galaxies, including galaxies that do not show young ages and high
metallicities from the Lick analysis (such as IC 4329 and NGC 4936),
is indicative that this component may be an artifact of the
fit. In particular, it might be an artifact due to the solar
[$\alpha$/Fe] of the models. In other words, the fit is skewed to
these very metal-rich intermediate age populations to make up for the
models being at solar abundances while the galaxies are known to have
[$\alpha$/Fe] $>$ 0. For this to be the case, however, the very metal
rich P{\'E}GASE-HR models should have super-solar [$\alpha$/Fe] but
the very metal rich Milky Way stars upon which the models are based
will have close to solar [$\alpha$/Fe] \citep{thomas03}. Therefore, it
is unclear if this could be the case. Fitting the galaxies with a
different set of SSP model spectra could decide this matter but, at
[Fe/H] $\sim$ 0.7, the subpopulation is out of the parameter space of
other models and certainly close to the edge of the parameter space of
the {\'E}lodie library upon which the P{\'E}GASE-HR models are
built. With that caveat, we still tried fitting the galaxy spectra
with the differential models of \citet{walcher09} which use the
theoretical models of \citet{coelho07} to alter the model spectra of
\citet{vazdekis10} to account for non-solar [$\alpha$/Fe]. Fitting our
galaxies with these model spectra reproduced the high [$\alpha$/Fe]
determined from the Lick indices but did not reproduce the
intermediate age subpopulation. This might be due to the limited
parameter space of these models as they extended only to [Fe/H] =
+0.2. To determine whether these subpopulations are real would require
further measurement. Further measurements could include using spectra with a wider wavelength range which might provide more stable fits with the models or model spectra with a wider range in parameters when such models are created. The $\sim$500Myr subpopulation in NGC 5670, on the
other hand, probably is real as it explains the very young
SSP-equivalent ages and is found only in NGC 5670. This subpopulation
could date the most recent merger episode and, most likely, the shells
to $\sim$500 Myr ago. This could be used as a constraint for simulations of
shell formation. If the $\sim$4-5Gyr old populations are real, they
could be used as well.

\subsubsection{The age of the NGC 5670 shell system}

It is possible to use the observed shell distribution of NGC 5670 to obtain an
independent estimate of the merger age by simulating the shell
structure semi-analytically \citep{bilek14}. Strictly speaking, this
estimate is only valid for a minor merger scenario whereby the
accretion is along the major axis. Since this may not be the case, we
seek to use this method to obtain an order of magnitude estimate for
the shell ages only.

In this scenario, shells are formed by stars on radial orbits that are
near their apocenters. The time required to make a shell is given by:
\begin{equation}
t = (n + 1/2)P(r_{a})
\end{equation}
where $P(r_{a}$) is the period of stars that reach their
apocenters at a radius of $r_a$. 
The shell number $n$ is obtained by counting from the outermost shell to the
innermost. Note that the shell with $n=0$ consists of stars reaching their first apocenters and may or may not be visible \citep{bilek14}. 
The shell's position can be expressed through the conservation of energy as:
\begin{equation}
\phi(r_{s}) + \frac{1}{2}v_{s}^{2}= \phi(r_{a})
\end{equation}
where $r_{s}$ is the shell's position, $v_{s}$ is the shell phase
velocity, and $\phi$(r) is the gravitational potential at position
$r$. Following \citet{bilek14}, we approximate the shell phase velocity
as:
\begin{equation}
v_{s} \approx \frac{1}{(n+1/2)\frac{dP}{dr}(r_{a})}
\end{equation}

From the expressions in \citet{cap13}, we can estimate the stellar
mass of NGC 5670 as $4.7\times10^{10}$$M_{\odot}$, based on the velocity
dispersion out to $1\ R_{\mathrm{eff}}$ of $\sim 110\ \kms$. 

\begin{figure}
 \includegraphics[scale=0.32]{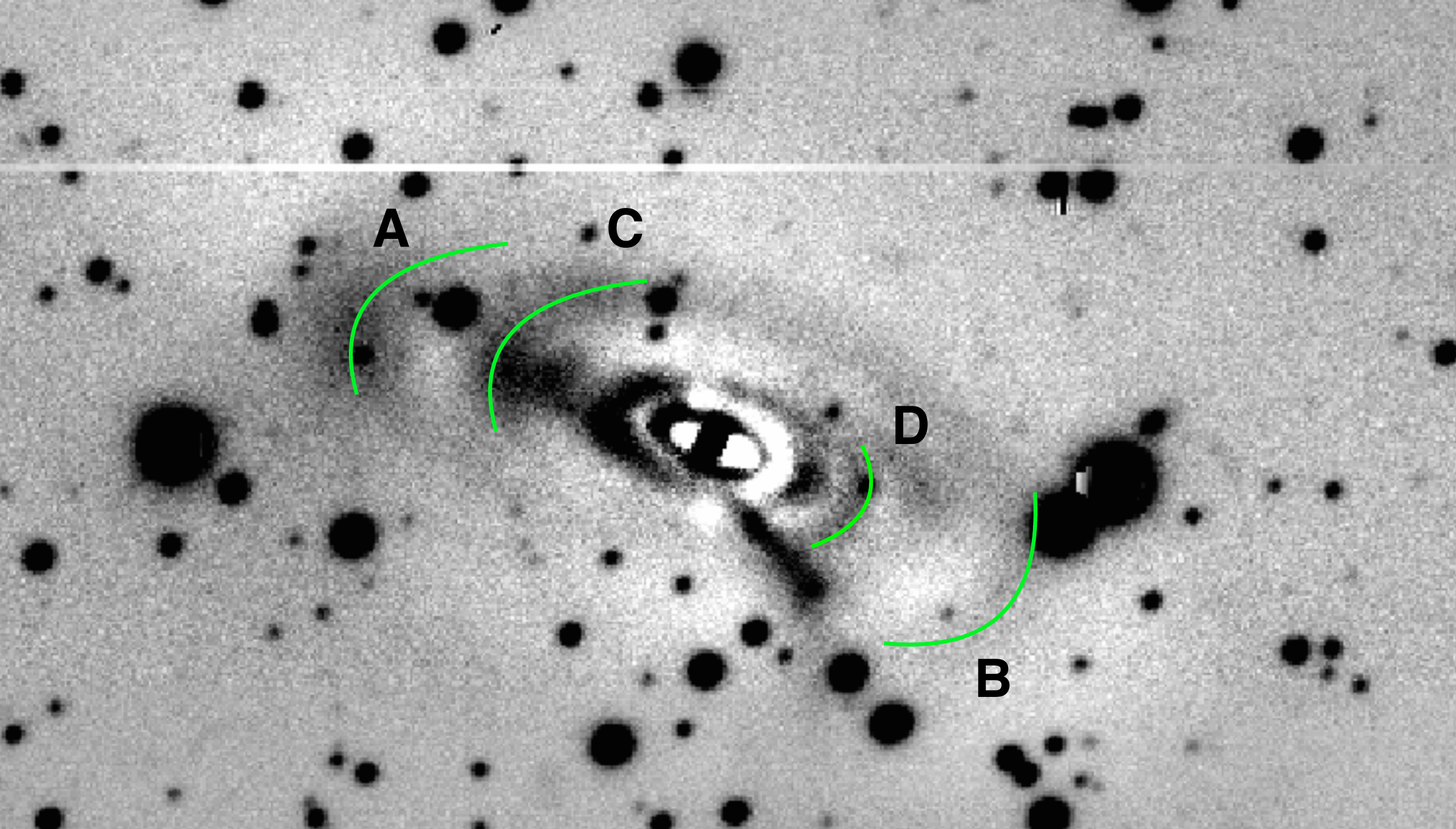}
\caption{DECam image of NGC 5670 with the host galaxy fitted and subtracted out. The four outermost shells are labeled A-D.}
\label{fig:shells5670}
\end{figure}

\begin{table}
\centering
\begin{tabular}{||c | c c c c ||} 
 \hline
 M ($M_{\odot}$)  & Label & n & r ('') & t (Myr) \\ [0.5ex] 
 \hline
 4.7x10$^{10}$ & A & 1 & 35.7 & 170  \\ 
  & B & 2 & 31.6 & 235  \\
  & C & 3 & 20.4 & 193  \\
  & D & 4 & 15.1 & 173  \\

\hline
 2.4x10$^{11}$ & A & 1 & 35.7 & 76  \\ 
  & B & 2 & 31.6 & 105  \\
  & C & 3 & 20.4 & 86  \\
  & D & 4 & 15.1 & 77  \\[1ex] 
 \hline
\end{tabular}
\caption{Time scales required for shell formation for each of the shells observed. These are calculated for two masses of NGC 5670: one with $M/L=1$ and another with $M/L=5$.}
\label{table:1}
\end{table}

To find the shells, we use an archival DECam image
of NGC 5670. Using GALFIT \citep{peng02} we find the galaxy is
reasonably well fit with a de Vaucouleurs profile combined with a Sersic profile after masking out the shells. Figure \ref{fig:shells5670} shows the galaxy after the underlying profile is subtracted out. We note that the location of the shells is not changed if the galaxy is fit with a single de Vaucouleurs profile or only a single Sersic profile. Several shells are visible in addition to a tidal stream feature seen towards the bottom right of the galaxy. We label what appear to be the four outermost shells A-D. The shell structure appears to be somewhat complex with some shells completely surrounding the galaxy. There also appears to be an incomplete shell between shells D and B in figure \ref{fig:shells5670}. This is all suggestive that the shells likely did not form in an idealistic minor merger along the galaxy's major axis. However, we will continue to use this as a model to get an order of magnitude estimate of the age of the system.

To obtain the shape of the potential, we deproject a de
Vaucouleurs profile to obtain a density profile.  We adopted an oblate
rotator configuration with a minor- to major- axis ratio of $0.46$
(Hau, et al., in preparation).  Using equations 6 and 7, we estimate
$r_{a}$, and using equation 5 we calculate the age required to produce a shell of a given radius. The ages required to produce shells A-D given their radii from the DECam image are listed in table 3. To account for a different galaxy mass, we
also calculate the shell ages if the mass was 5 times
higher. Note, we assume that the outermost shell in figure \ref{fig:shells5670} is the shell with $n=1$. If it was actually the $n=0$ shell, the ages in table 3 would be roughly half as much.

For the case with $M =  4.7$x$10^{10} M_{\odot}$, ages are all consistent with a merger roughly $200$ Myr ago whereas with $M =  2.4$x$10^{11} M_{\odot}$, ages are consistent with a merger roughly $100$ Myr ago. Given the uncertainty in our value for the mass of NGC 5670, these ages are quite close to
the age of the young subpopulation shown in figure~\ref{fig:massweight5670}. 
Therefore it is
reasonable to conclude that a recent gas-rich merger event gave rise
to both the young subpopulation as well as the shell system in NGC
5670.

\section{Conclusions}
\label{sec:conclusions} 

In  this  work  we  have  presented  long  slit  spectra  of  9  shell
galaxies. We analyze the data through two methods to determine the stellar population parameters of these galaxies. First, we use the Lick indices
and second, we perform full spectrum fitting. While we find a fair agreement between both
methods particularly in metallicity, discrepancies are found in the age for some galaxies.
We argue that this is likely  due  to  the presence  of  young stellar  populations
affecting the SSP-equivalent parameters derived from the Lick indices
and to the age-metallicity degeneracy that plagues index use.  For these reasons, we rely
on the full-spectrum fitting for much of the analysis.

Throughout our analysis, we find many indicators that the shell galaxies have undergone past interaction events. Many of the shell galaxies in our sample appear to have lower central $\mathrm{Mg}_{2}$ index values than that expected from the \citet{bender93} relation based on non-shell galaxies. We argue that this trend is caused by recent merging events. Several of the shell galaxies in our sample exhibit ionized gas emission and young subpopulations of stars, which we take as more evidence of recent interactions. In particular NGC 5670 shows a pronounced departure from the \citet{bender93} relation relation due to the presence of a young stellar sub-population.

Our shell galaxy sample shows a relation between central metallicity and velocity dispersion that is statistically consistent with a previous sample of non-shell galaxies but appears to be slightly displaced towards lower metallicity which we again argue that, if real, is due to recent interactions.

With  the derived population parameters, we calculate the metallicity gradients for the
sample. We find negative metallicity gradients in all 9 galaxies, with an average metallicity gradient of $grad([Z/H]) = -0.16 \pm
0.10$ dex per decade in radius. Our galaxies exhibit relatively
shallow metallicity gradients which compare well with those predicted
for major mergers by \citet{kobayashi04}. 
This result and the evidence for recent merging events mentioned above seem to rule out the weak interaction scenario as the cause of the shells and support the picture that the shells form from merging events. Unfortunately, it is difficult to say whether the shells formed in major merging events or in separate minor, accretion events. More detailed measurements of shell locations could be combined with simple simulations and the ages of subpopulations to try to further distinguish between major and minor mergers.

As mentioned above, we find evidence for young starburst
subpopulations in around half of our galaxy sample with ages ranging from $\sim500$ Myr to
$\sim4$ Gyr. These are suggestive of the age of a previous merging episode
and possibly the age of the shell system. However, as discussed in $\S4.3$, they may also be
artifacts of the fitting procedure. 

For NGC 5670, which has the most prominent young subpopulation, we
find that the observed shell structure can be formed in $\sim200$ Myr which is
consistent with the age of the starburst population. Therefore, at
least in NGC 5670, a single merger event may have triggered a
starburst as well as forming the shell system, consistent with a
wet-merger scenario.

\section*{Acknowledgments}
This project was conducted within the framework of the CTIO REU program, which is supported by the National Science Foundation under grant AST-1062976.

\label{lastpage}

\end{document}